\documentclass[twocolumn,aps,showpacs,groupedaddress,floatfix,pre]{revtex4-1}
\usepackage{amsmath}
\usepackage{graphicx}
\usepackage{dcolumn}
\usepackage{bm}
\usepackage{verbatim}
\usepackage{color}
\usepackage{hyperref}
\usepackage{ulem}
\usepackage{float}
\begin{document}
\setlength{\parskip}{0mm}

\title{Connecting Relaxation Time to a Dynamical Length Scale in Athermal Active Glass Formers}
\author{Dipanwita Ghoshal}
\affiliation{Department of Physics, Indian Institute of Technology Madras, Chennai, Tamil Nadu 600036, India}

\author{Ashwin Joy}
\email{ashwin@physics.iitm.ac.in}
\affiliation{Department of Physics, Indian Institute of Technology Madras, Chennai, Tamil Nadu 600036, India}
\date{\today}

\begin{abstract}
  Supercooled liquids  display dynamics that are inherently heterogeneous in space. This essentially means that at temperatures below the melting point, particle dynamics in certain regions of the liquid can be orders of magnitude faster than other regions. Often dubbed as dynamical heterogeneity, this behavior has fascinated researchers involved in the study of glass transition, for over two decades. A fundamentally important question in all glass transition studies is whether one can connect the growing relaxation time to a concomitantly growing length scale. In this paper, we go beyond the realm of ordinary glass forming liquids and study the origin of a growing dynamical length scale $\xi$ in a self propelled ``active'' glass former. This length scale which is constructed using structural correlations agrees well with the average size of the clusters of slow moving particles that are formed as the liquid becomes spatially heterogeneous. We further report that as the effective temperature $T_\text{eff}$ is lowered, the concomitantly growing $\alpha$- relaxation time exhibits a simple scaling law, $\tau_\alpha \sim \text{exp} (\xi \mu / T_{\text{eff}})$, with $\mu$ as an effective chemical potential, and $\xi \mu$ as the growing free energy barrier for cluster rearrangements. The findings of our study are valid over a wide duration of self-propulsion, and hence could be very useful in understanding the slow dynamics of a generic active liquid such as an active colloidal suspension, or a self propelled granular medium, to mention a few. 

\end{abstract}


\maketitle

\section{\label{intro}Introduction}

It is well known that liquids near a glass transition display a growing length scale \cite{ediger2000spatially,richert2002heterogeneous,donati1999spatial,berthier2004time,karmakar2009growing,flenner2011analysis} that is normally associated with the concomitant sluggish dynamics. This growing length scale in a typical glass former often characterizes the spatial extent of dynamic heterogeneity and the literature on this subject is quite extensive \cite{glotzer2000time,richert2002heterogeneous,andersen2005molecular}. The aforementioned heterogeneity in space and time is ubiquitous in nature and also observed in active systems having an internal source of energy, ranging from biological cells \cite{angelini2011glass} and tissues \cite{schoetz2013glassy}, to even ant colonies \cite{gravish2015glass}. The visual identification of dynamic heterogeneity in a glass forming liquid at low temperatures often requires a definition of a cluster of particles moving slower than the surrounding medium. The average size of these clusters which can be treated as a growing length scale, tends to increase as one approaches the glass transition, and the cooperative movement of these large clusters results in the observed sluggish dynamics. There exist numerous reports on the study of growing length scales in passive glass forming liquids, that cover extensive simulations \cite{donati1999spatial,lavcevic2003spatially,whitelam2004dynamic,berthier2004time,stein2008scaling,karmakar2009growing,karmakar2010analysis}, experiments \cite{berthier2005direct,dalle2007spatial,crauste2010evidence} and theoretical investigations \cite{biroli2004diverging, biroli2006inhomogeneous, berthier2007spontaneous, berthier2007spontaneous, szamel2008divergent, szamel2010diverging}. However, similar explorations of a growing length scale in active glass formers remain limited in scope. Here we study the dynamic heterogeneity in a system of self-propelled particles and explore various growing length scales that are concomitant with sluggish dynamics near the glass transition. The active liquid in our study is represented by a minimal Ornstein-Uhlenbeck (OU) model in which the activity is completely described by only two parameters, the persistence time $\tau_p$ and the effective temperature $T_{\text{eff}}$. The model can be used to study active colloidal suspensions, passive tracers in bacterial baths, and self-propelled granular media, to mention a few. In what follows, we provide a brief background covering some results that are relevant to our study.

It is well established that the four point structure factor is maximum around the $\alpha$- relaxation time ($\tau_{\alpha}$) for typical glass forming systems \cite{berthier2004time, karmakar2009growing}, and can be profitably used to extract a dynamical length scale $(\xi)$. In this paper, we have extracted $\xi$, and two other length scales, namely the microscopic hexatic length scale $\xi_6$ and the diffusion length scale $\sqrt{D \tau_\alpha}$\cite{lerner2009statistical}. We also find that at lower temperatures, the dynamical length scale $\xi$ decouples from all the other length scales examined here- an observation made earlier in some works on passive glass formers \cite{tah2018glass, kawasaki2010structural, kawasaki2011structural}. The dynamic length scale $\xi$ extracted from long range structural correlations can be connected to the concomitant growing $\alpha$- relaxation time scale $\tau_\alpha$, through a simple scaling law, $\tau_\alpha \sim \text{exp} (\xi \mu / T_{\text{eff}})$, that is similar to predictions from the random first order transition theory (RFOT)\cite{sastry1998signatures, karmakar2014growing, kirkpatrick2015colloquium} in two dimensions. Our results are valid over three decades of persistence time and at every $\tau_p, T_{\text{eff}}$ state point, the size of the slow moving clusters agrees well with the dynamic length scale $\xi$. We conclude the paper with the role of persistence time $\tau_p$ on the chemical potential $\mu$ which is essentially an energy budget per unit length  to form clusters of slow moving particles.

The paper is organized into the following sections. Section (\ref{intro}) discusses the introduction and necessary background of the subject, section (\ref{model}) presents the minimal model used in our work to describe an athermal active glass forming liquid, section (\ref{results}), and finally in section (\ref{summary}), we summarize our work and present interesting future directions resulting from our work. We will now discuss the numerical model used in our work.

\section{\label{model} Minimal Model}
There is enough evidence to suggest that  fluids composed of active particles can exhibit glassy dynamics at sufficiently high particulate densities, notable mentions being experiments on migrating cells \cite{angelini2011glass} and  embryonic tissues \cite{schoetz2013glassy}, as well as numerical simulations \cite{flenner2016nonequilibrium}. There have been some recent reports on the influence of activity on the sluggish dynamics in glass forming systems \cite{mandal2016active,flenner2016nonequilibrium,mandal2017glassy,giavazzi2018flocking,lei2019nonequilibrium} but the results depend qualitatively on the presence of thermal noise in the system. In our work, we use an athermal model of self-propelled particles introduced earlier \cite{flenner2016nonequilibrium,fodor2016far,donado2017brownian,das2018confined}. The model is deemed athermal as there is no explicit presence of thermal noise term in the position update of the particles. The governing equations for the i$^{th}$ particle with mass $m$ reads as,
\begin{eqnarray}
  \bm{\dot r}_{i} &=& \frac{1}{m\gamma}\biggl[- \sum_{j\neq i} \bm \nabla_i \phi (r_{ij}) + \bm f_i\biggr]\nonumber\\
  \bm{\dot f}_{i} &=& \frac{1}{\tau_p}\biggl[-\bm f_i + \sqrt{2m\gamma k_{B}T_{\text{eff}}}\;\boldsymbol{\eta}_{i}\biggr]
\end{eqnarray}
Put simply, the dynamics is overdamped with $m\gamma$ as the friction
coefficient and falls under the Ornstein-Uhlenbeck type of stochastic process
that has been used to model athermal active systems.  The self-propulsion force
$\bm f_i$ is a colored noise with an
exponentially decaying auto-correlation function given by\\
    \begin{equation}
      \langle \bm f_{i}(t) \bm f_{j}(t^{\prime}) \rangle = \biggl(\frac{m\gamma k_B T_{\text{eff}}}{\tau_{p}}\biggr) \bm I \delta_{ij}e^{-|t-t^{\prime}|/\tau_{p}},
   \end{equation}
   where $\bm I$ is the identity matrix. We take $\bm \eta_i$ to be a Gaussian white noise of zero mean and variance
   \begin{equation}
     \langle \bm \eta_{i}(t) \bm \eta_{j}(t^{\prime}) \rangle = \bm I \delta_{ij} \delta (t-t') 
     \label{}
   \end{equation}
   The particles interact through the Lennard-Jones potential, 
   \begin{equation}
     \phi(r_{ij}) = 4\epsilon_{ij} \biggl[ \biggl(\frac{\sigma_{ij}}{r_{ij}}\biggr)^{12} - \biggl(\frac{\sigma_{ij}}{r_{ij}}\biggr)^6\biggr]
     \label{}
   \end{equation}
We have taken a Kob-Andersen (KA) binary glass former where the ratio of large to small particles is kept 80:20. We took $\epsilon_{LL}$, $\sigma_{LL}$ and $\sqrt{m \sigma_{LL}^2 / \epsilon_{LL}}$ as the units of energy, length and time, respectively. In these units, the potential parameters become $\epsilon_{SS}$ = 0.5, $\epsilon_{LS}$ = 1.5, $\sigma_{SS}$ = 0.88, and $\sigma_{LS}$ = 0.80. Since our primary objective was to extract a growing length scale through long wavelength structural correlations, we have employed a two-dimensional (2D) periodic box with 10000 particles that gives us a large enough system size required for our work. To assert the robustness of our results, we perform our study over a wide range of persistence times $\tau_p$, ranging from 0.0002 which corresponds to the limit of equilibrium Brownian dynamics (BD), to all the way up to 0.1, which is far from equilibrium. Below we discuss our results.\\

\section{\label{results} Results and Discussion}
 \subsection{Dynamic Length Scale ($\xi$)}
\begin{figure}[h]
     \centering
     \includegraphics[width=0.85\textwidth,keepaspectratio]{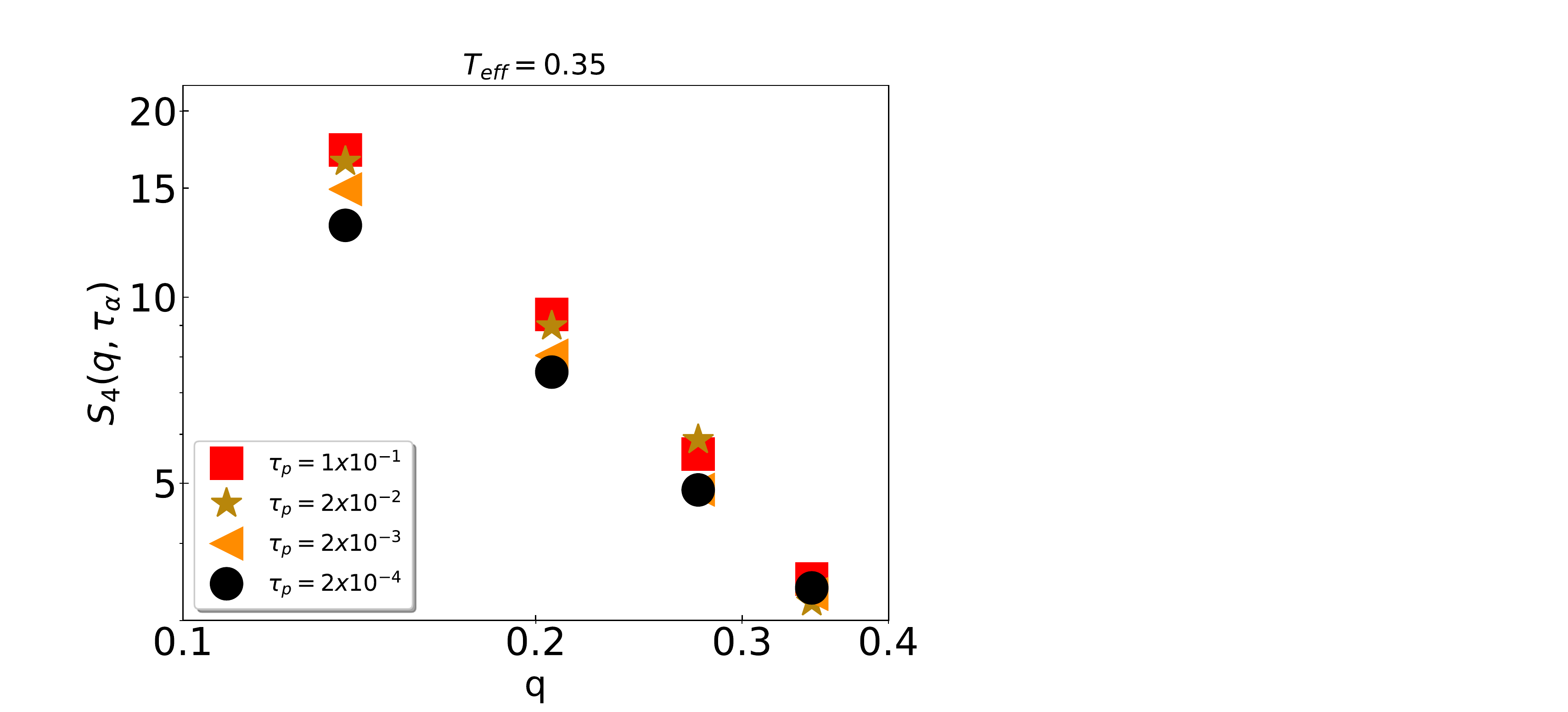}
     \caption{Four point structure factor as a function of wavevector at a typical low temperature. The long wavelength ($q \rightarrow 0$) collective behavior is amplified by increasing the persistence time $\tau_p$. One can therefore expect the formation of larger clusters with increasing $\tau_p$ at any given effective temperature.}
     \label{S4-vs-q}
   \end{figure}

\begin{figure}[h]
  \centering
\includegraphics[width=0.85\textwidth,keepaspectratio]{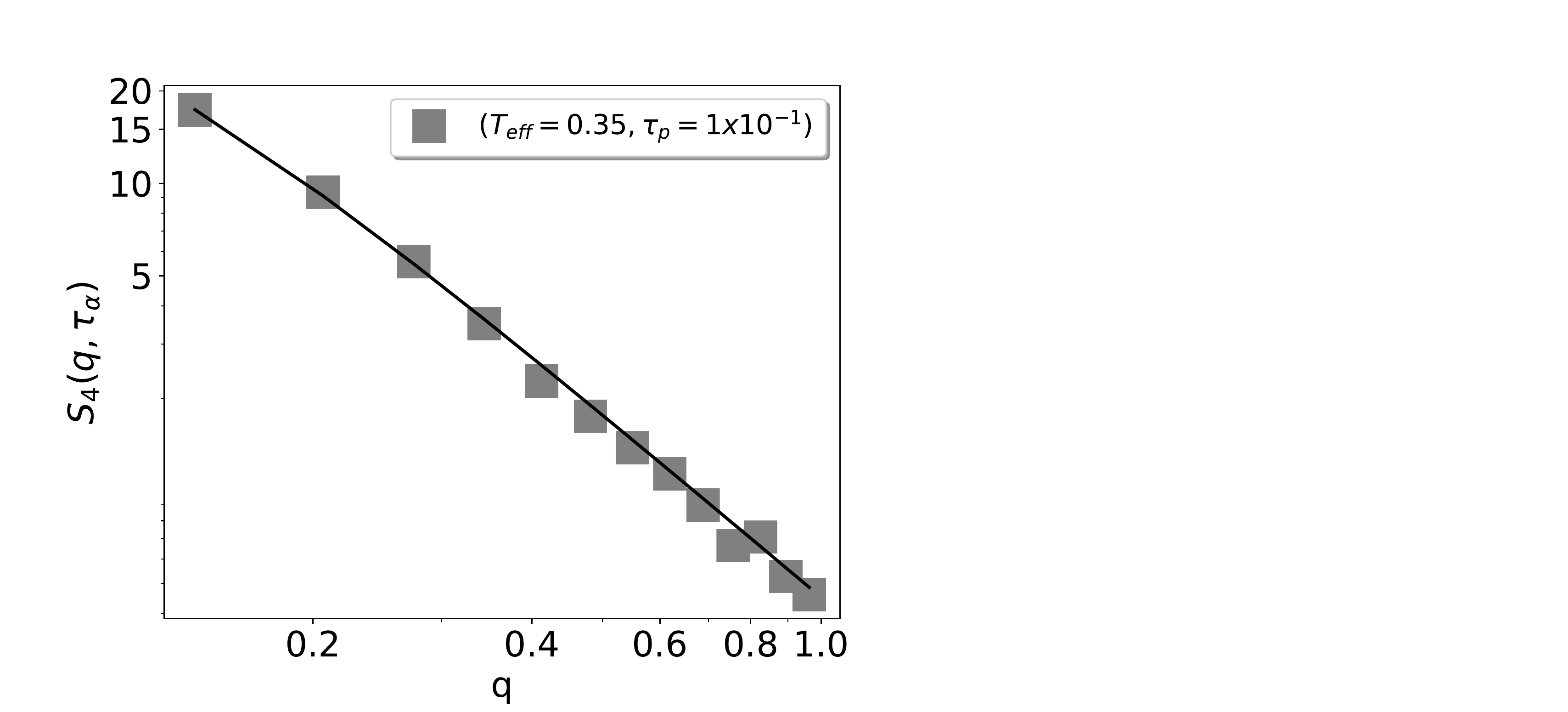}
\caption{Our data on $S_{4}(q,\tau_{\alpha})$ for a typical ($\tau_p, T_{\text{eff}}$) pair. Solid line is a fit to the OZ relationship mentioned in Eq. (\ref{OZ-eq}), and the extracted dynamical length scale $\xi$ is plotted in Fig. \ref{xi-fig}.}
\label{OZ-fit}
\end{figure}

\begin{figure}[h]
\centering
\includegraphics[width=0.85\textwidth,keepaspectratio]{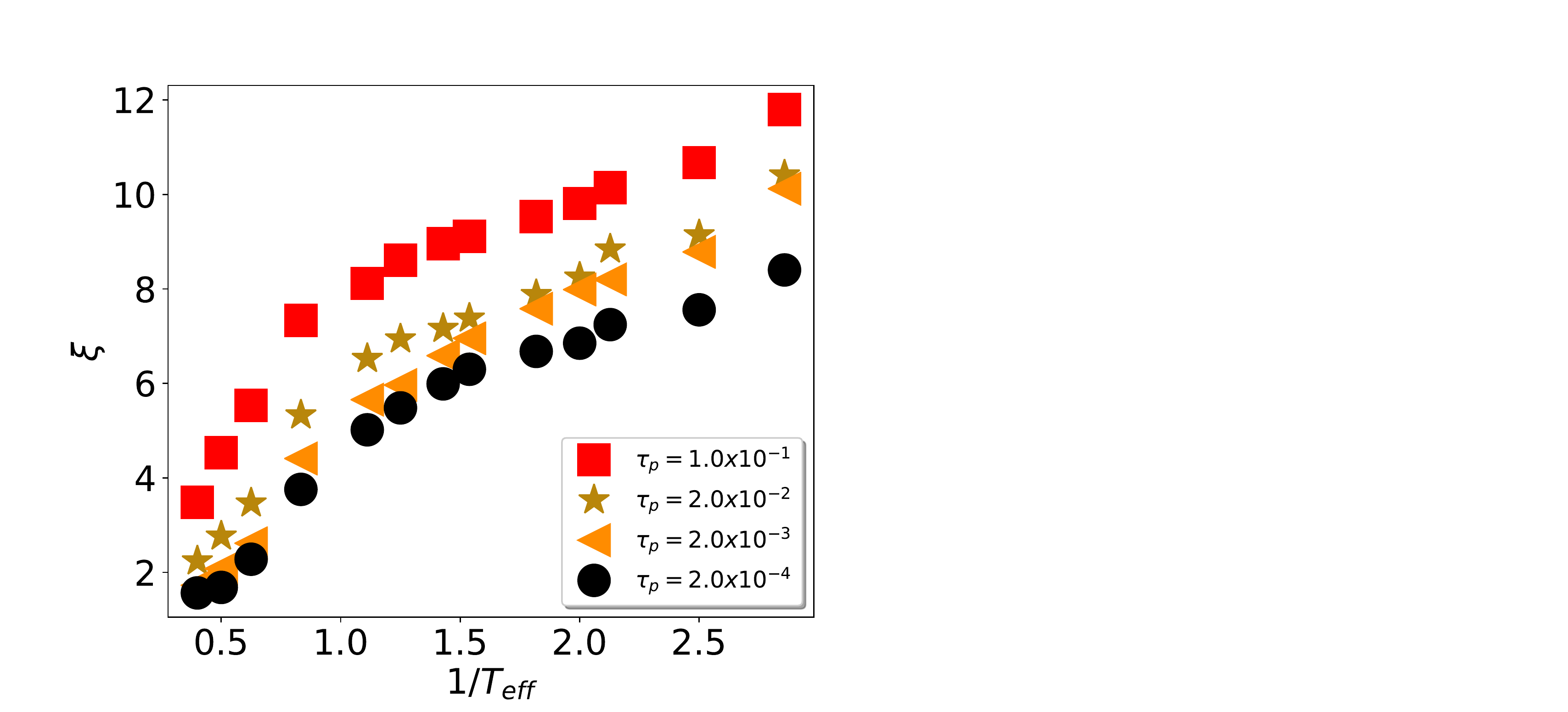}
\caption{At any effective temperature, the length scale $\xi$ increases with the persistence time $\tau_p$, and therefore results in a higher degree of spatial heterogeneity. This behavior manifests in the emergence of larger clusters of slow moving particles at higher values of persistence times at any specific effective temperature, see Fig. \ref{cluster-fig}}
\label{xi-fig}
\end{figure}

\begin{figure*}[h]
  \includegraphics[width=0.85\textwidth,keepaspectratio]{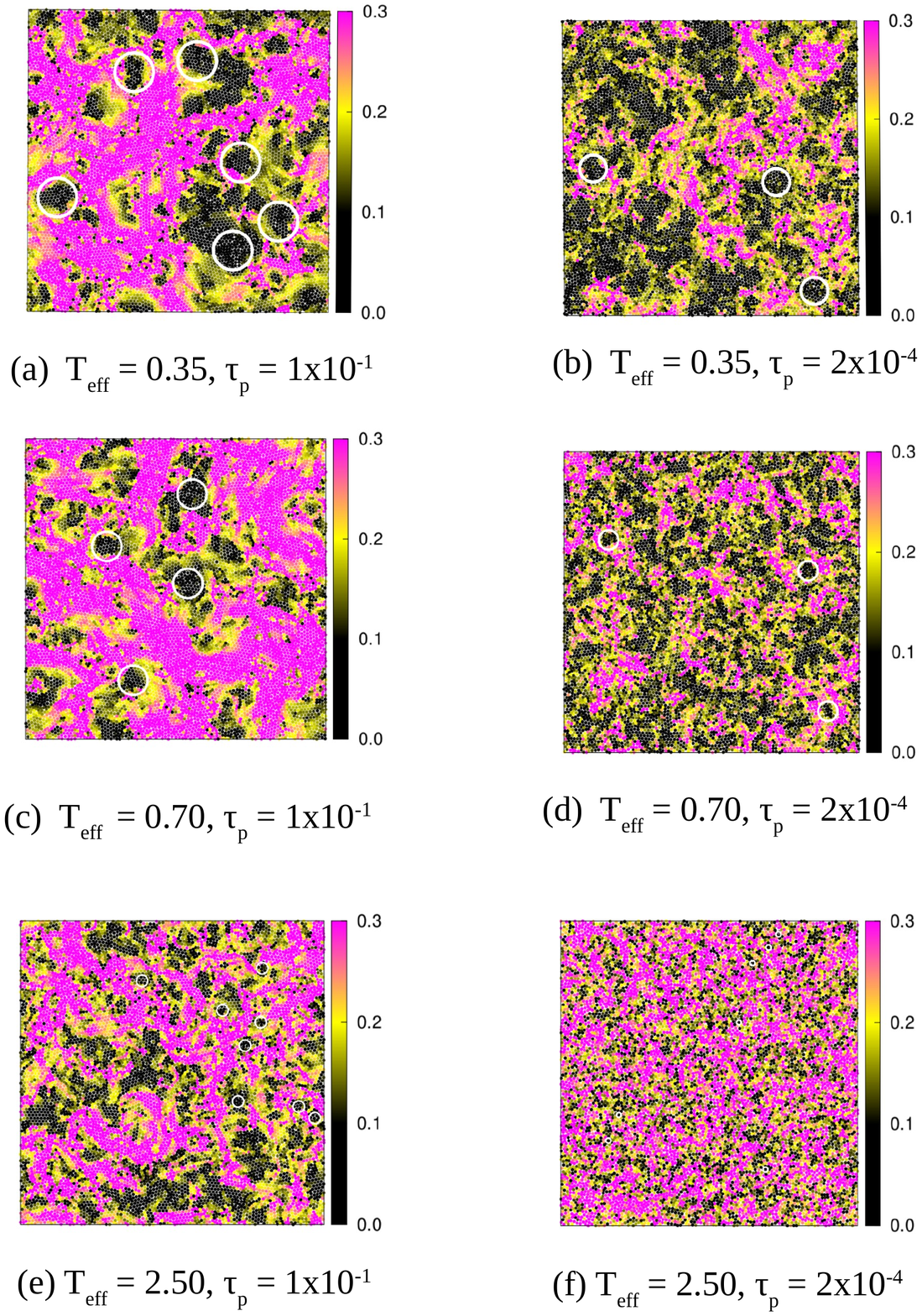}
	\caption{Snapshots of dynamical clusters for various $T_{\text{eff}}$ and $\tau_{p}$ from ((a)-(f)). Color bar indicates  magnitude of particle displacements recorded over a specified time which is set to be of the order of $\tau_\alpha$.  Encircled regions are plotted with $\xi$ extracted from OZ relation and it is evident that the cluster sizes are consistent with the extracted values of $\xi$.}
\label{cluster-fig}
\end{figure*}

 To extract the dynamic length scale $\xi$, we will start from the four-point structure factor, 
  \begin{equation}
     S_{4}(q; t) = \frac{1}{N}\biggl(\langle Q(\bm q; t) Q(- \bm q; t)\rangle - |\langle Q(\bm q; t) \rangle|^2  \biggr)
  \end{equation}
  where $Q(\bm q; t)$ is defined as the Fourier transform,
  \begin{equation}
     Q(\bm q; t) = \sum_n w_n(t) \text{exp}\; [-i \bm q \cdot r_n(0)],
  \end{equation}
  of the microscopic overlap function,
  \begin{equation}
  w_{n}(t) = \Theta [a-|\bm r_{n}(t) - \bm r_{n}(0)| ]
  \end{equation}
  Here $\Theta(x)$ is the Heaviside's step function and $\bm r_n(t)$ is the position of $n^{th}$ particle at time $t$. This function filters out the particles that do not move farther than a specified distance $a$, during a time interval $t$. We take $a = 0.3$, which roughly corresponds to the plateau value of the mean squared displacement (MSD) at all temperatures. Denoting the $\alpha$-relaxation time, $\tau_{\alpha}$ to be the time where the self-intermediate scattering function falls $1/e$ of its initial value, we plot our data on $S_4(q; \tau_\alpha)$ at a typical low effective temperature, in Fig. \ref{S4-vs-q}. It is evident from the figure that at any given effective temperature, the long wavelength ($q \rightarrow 0$) response increases with the persistence time $\tau_p$, suggesting us that the dynamical length scale $\xi$  must increase with the persistence time $\tau_p$. To extract $\xi$ at any fixed pair of ($\tau_p, T_{\text{eff}}$), we fit our $S_{4}(q; \tau_{\alpha})$ data to the following Ornstein-Zernike (OZ) relationship,
  \begin{equation}
  S_{4}(q; \tau_{\alpha}) = \frac{A}{1 + (q\xi)^{2}}.
  \label{OZ-eq}
\end{equation}
where $A = \lim_{q\rightarrow 0} S_{4}(q; \tau_{\alpha})$ and $\xi$ are fitting parameters for the chosen ($\tau_p, T_{\text{eff}}$) pair, see Fig. \ref{OZ-fit}. The extracted dynamical length scale $\xi$ is plotted in Fig. \ref{xi-fig} where it is evident that at all temperatures, the effect of $\tau_p$ is to increase the length scale $\xi$. This explains the formation of larger clusters of slow moving particles as $\tau_p$ is lifted at any given $T_{\text{eff}}$, see Fig. \ref{cluster-fig}.
\noindent

The method used here relies on the interplay between dynamics and structure near the glass transition, there are however, alternative approaches in the context of passive liquids, most notably through inherent structures that are essentially the minima in the rugged potential energy landscape of glass forming liquids \cite{goldstein1969viscous,stillinger1982hidden,sastry1998signatures,debenedetti2001supercooled,sciortino2005potential}. We believe our approach is simpler and has the potential to yield a deep insight into the sluggish dynamics of active glass formers, through the concept of a chemical potential $\mu$ (described later) that depends on the persistence time $\tau_p$. In what follows, we will show that an increase in the persistence time leads to a larger free energy barrier for cluster rearrangements, directly leading to the concomitantly growing time scale $\tau_\alpha$. The reader should note that the average size of the clusters shown as white circles in the Fig. \ref{cluster-fig}, agrees well with the length scale $\xi$ extracted earlier from the four point structure factor. The cluster themselves are identified by particles that traveled a distance less than $a=0.3$ (slow moving) in a pre-specified time interval, $\tau_\alpha$. As is evident from the figure, at higher effective temperatures $T_{\text{eff}}$ and lower persistence times $\tau_p$, the system becomes increasingly homogeneous with the cluster size becoming very smaller. On the other hand, decreasing either $T_{\text{eff}}$ or increasing $\tau_p$ leads to an increase in cluster size, and therefore spatial heterogeneity. The largest clusters in our study appear in the state $T_{\text{eff}} = 0.35$ and $\tau_p = 0.1$. Below we present a simple scaling relationship that connects this growing length scale $\xi$ to the concomitantly growing time scale $\tau_\alpha$ that remains valid over a wide range of persistence times.

\subsection{Chemical Potential ($\mu$) as an Energy Barrier for Cluster Relaxation}
\begin{figure}[h]
  \centering
\includegraphics[width=0.85\textwidth,keepaspectratio]{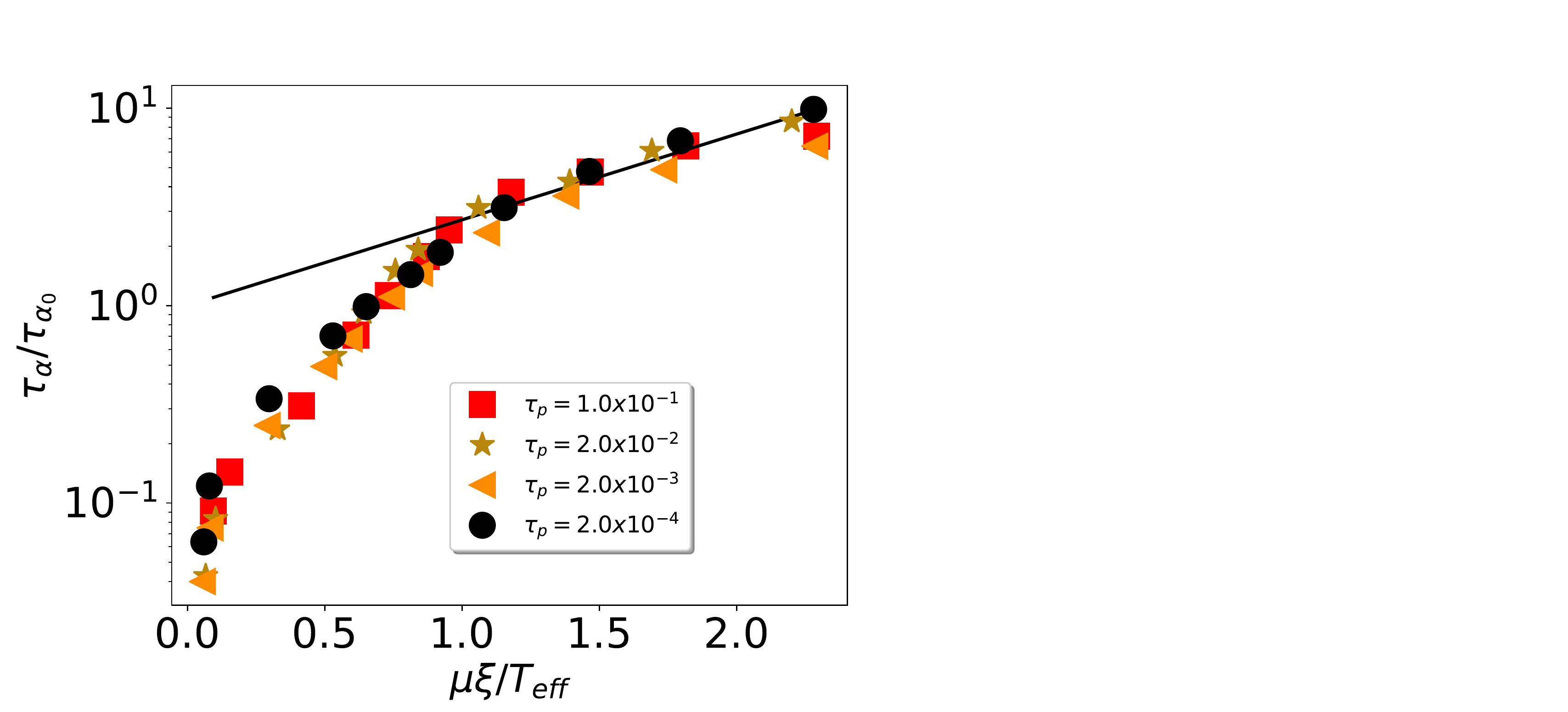}
\caption{Demonstration of data collapse for the scaled time $\tau_\alpha/\tau_{\alpha 0}$ vs. scaled free energy $\mu \xi/T_{\text{eff}}$ achieved over three decades of $\tau_p$. The solid line is a fit to the scaling law in Eq. \ref{tau-xi-scaling}. Note that the relaxation time is predicted to have a true divergence only at $T_{\text{eff}} \rightarrow 0$. At very high temperatures, well defined clusters do not exist and the scaling law overestimates the relaxation times.} 
\label{xi-tau-collapse}
\end{figure}

If we consider $\mu$ as a chemical potential that needs to be scaled for the formation of a cluster of a certain size $\xi$, then 
we may assume a simple scaling formula for the relaxation (rearrangement) time of these clusters,
\begin{equation}
\tau_{\alpha}=\tau_{\alpha 0}\text{exp}\left[\frac{\mu\xi}{T_{\text{eff}}}\right]
\label{tau-xi-scaling}
\end{equation}
where $\mu$ and $\tau_{\alpha 0}$  are treated as fitting parameters at any
fixed $\tau_p$. The product $\mu \xi$ then refers to the growing free energy
barrier of cluster rearrangements as the effective temperature is lowered, at
any fixed $\tau_p$. The form of our scaling law is similar to the predictions
of the random first order transition theory (RFOT) in two dimensions
\cite{lerner2009statistical,kirkpatrick1989scaling,kirkpatrick2015colloquium,karmakar2014growing}.
Recently Nandi \textit{et al} extended RFOT in active liquids by capturing the
modulation of configurational entropy by  activity, to predict the relaxation
time scales in the active liquids \cite{nandi2018random}. Our work however,
presents a simplified picture that connects the growing dynamical length scale
to the $\alpha$- relaxation in active liquids through the concept of a growing
free energy barrier of cluster rearrangements. We are able to collapse the
entire simulation data through this picture by using just two fitting
parameters, namely $\mu$ and $\tau_{\alpha 0}$, and our work has therefore
great utility in experiments targeting dynamical heterogeneity in such systems.
In Fig.  \ref{xi-tau-collapse}, we show a data collapse to the scaling law
reported in Eq. \ref{tau-xi-scaling}, over three orders of persistence time.
The quality of the collapse is evident from this figure, particularly at lower
effective temperatures, and all the way to the lowest effective temperature
accessible in our simulations. It should be noticed that our scaling law
predicts a divergence of the relaxation time only as $T_{\text{eff}}
\rightarrow 0$. The overestimation by scaling law at high temperatures may be
attributed to the lack of well defined clusters in this regime as particles do
not move cooperatively. The high temperature limit, $\tau_{\alpha} /
\tau_{\alpha 0} \rightarrow 0 \;(T_{\text{eff}} \rightarrow \infty)$
nevertheless, is consistent with the expectations from a viable scaling theory. 

\begin{figure}[h]
\includegraphics[width=0.85\textwidth,keepaspectratio]{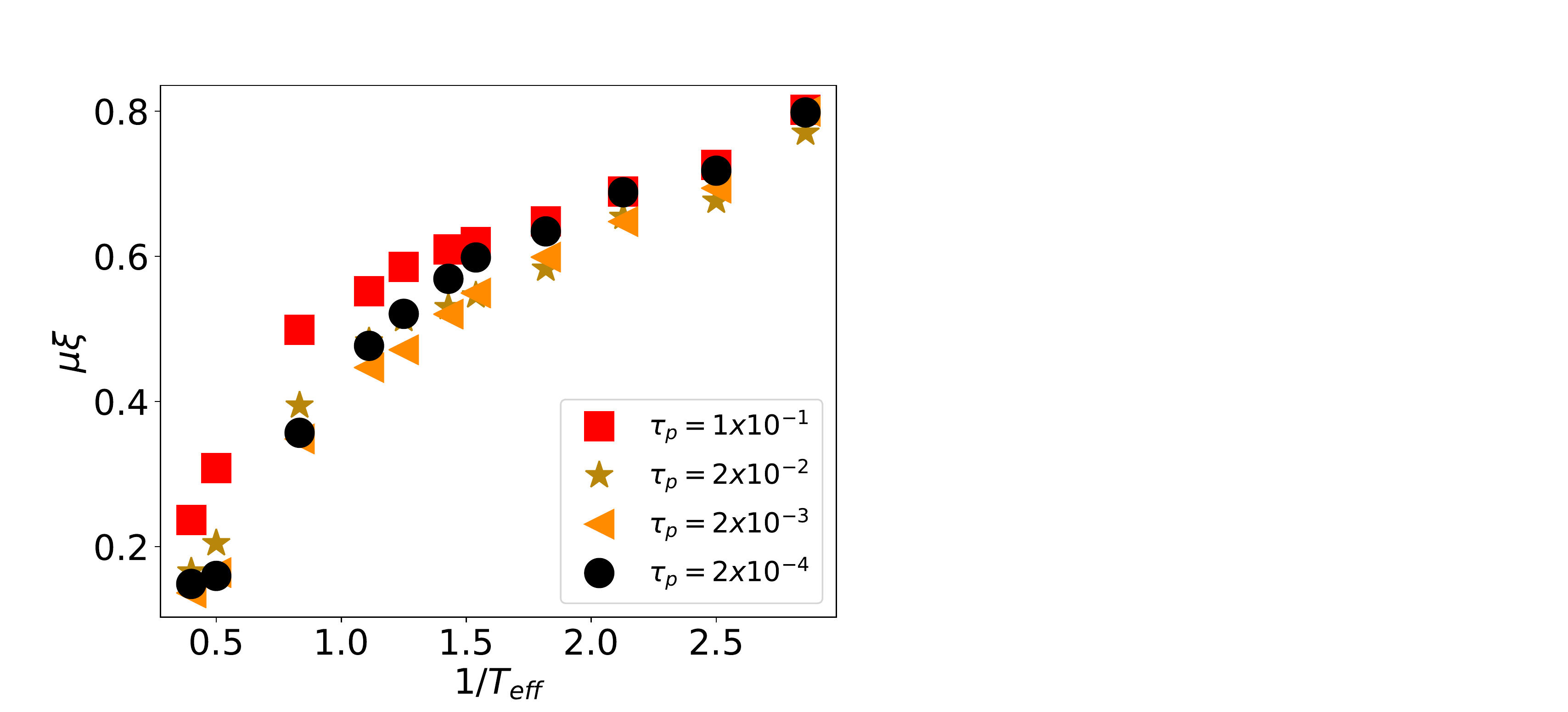}
\caption{Growing energy barrier $\mu\xi$ for cluster rearrangements as a function of effective temperature $T_{\text{eff}}$. The collapse is evident as $T_{\text{eff}} \rightarrow 0$, indicating the validity of the scaling law mentioned in \ref{tau-xi-scaling}.}
\label{muxi-vs-T}
\end{figure}

\begin{center}
\begin{table}[!htbp]
\begin{tabular}{|p{1cm}|p{1cm}|p{1cm}|}
\hline
$\tau_{p}$	& $\tau_{\alpha 0}$ &	$\mu$\\ 
\hline
0.1	& 15.0 &	0.068 \\
\hline
0.02 & 8.5&	0.074 \\ 
\hline
0.002 & 3.0 & 0.079\\ 
\hline
0.0002 & 1.9 & 0.095\\
\hline
\end{tabular}
\caption{Extracted values of $\tau_{\alpha 0}$ and $\mu$ for $\xi$ as fitting parameters. It is evident from the figure that increasing the persistence time lowers the chemical potential thereby facilitating the formation of large clusters of slow moving particles, see Fig. \ref{cluster-fig}.}
\label{mu-table}
\end{table}
\end{center}

To understand the effects of the persistence time $\tau_p$ on the chemical potential $\mu$ we provide a table \ref{cluster-fig} of our fitting parameters. It is evident from this table that increasing the persistence time or being farther from equilibrium results into a decrease in $\mu$, directly resulting in facilitating the formation of large clusters, see Fig. \ref{cluster-fig}. The energy barrier $\mu \xi$ associated with the clusters of size $\xi$ is seen to grow with $\frac{1}{T_{\text{eff}}}$, see Fig. \ref{muxi-vs-T}. The effect of persistence time has been beautifully captured by the energy barrier of cluster rearrangements. In what follows we shall explore some other relevant microscopic length scales in the active liquid that can be constructed from the underlying structure and particle dynamics.

\subsection{Hexatic Length Scale ($\xi_{6}$)}
\begin{figure}[h]
\includegraphics[width=0.85\textwidth,keepaspectratio]{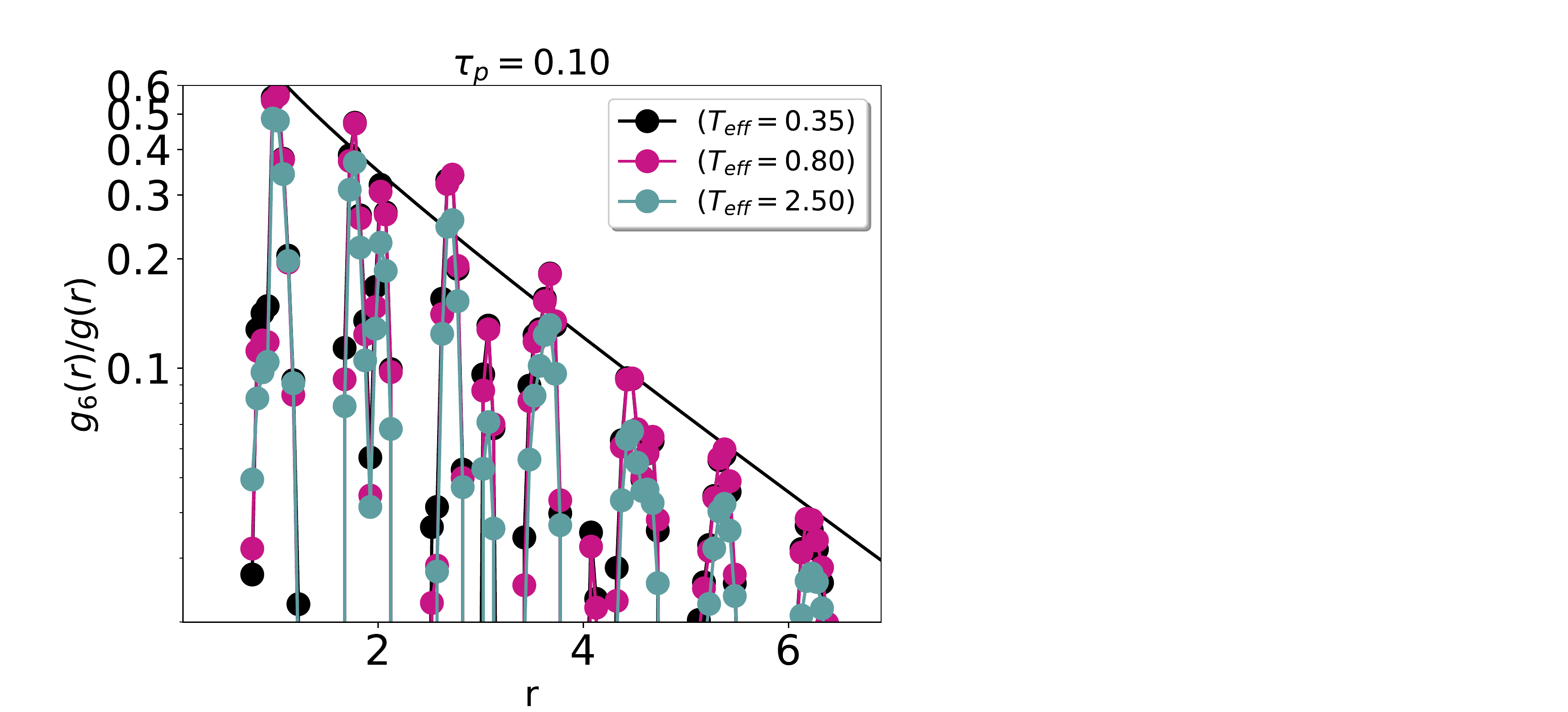}
\caption{A plot of $g_{6}(r)/g(r)$ at various values of effective temperature but a fixed persistence time $\tau_{p}= 0.10$. Solid lines that indicate a fit to the OZ relation \ref{g6-OZ-fit}, are used to extract a hexatic length scale $\xi_6$.} 
\label{g6-fit}
\end{figure}

\begin{figure}[h]
\includegraphics[width=0.85\textwidth,keepaspectratio]{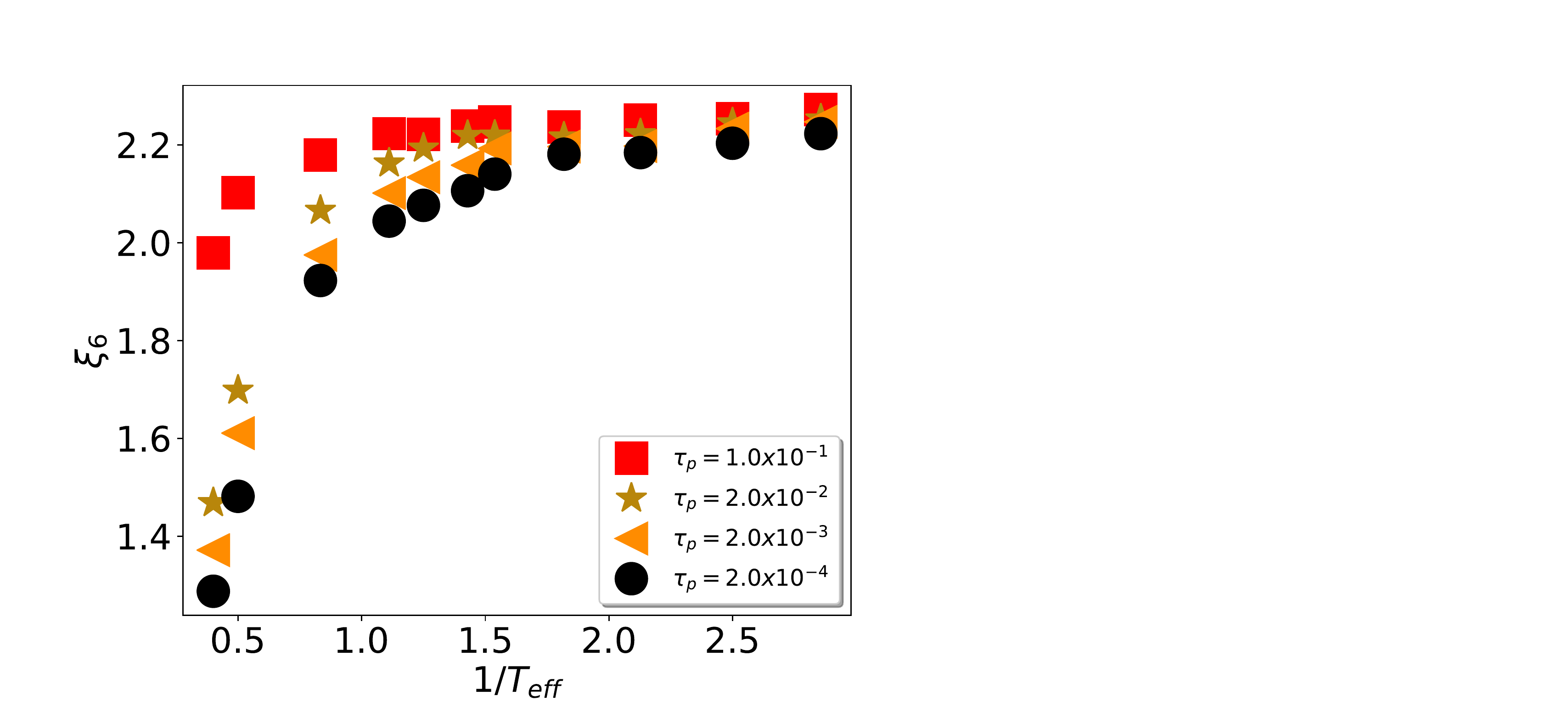}
\caption{At lower effective temperatures ($1/T_{\text{eff}} > 1.0$), the hexatic length scale $\xi_6$ grows very weakly. Moreover, the effect of $\tau_p$ on $\xi_6$ at these temperatures is also marginal. Therefore we are led to the conclusion that $\xi_6$ is not the dominant length scale that is connected to the growing relaxation time.} 
\label{xi6-vs-T}
\end{figure}

The hexatic length scale $\xi_6$ is a well known static length scale that has been  used to study medium range crystalline order (MRCO) in 2D passive glass forming liquids \cite{kawasaki2007correlation,tanaka2010critical,kawasaki2011structural,tah2018glass}.
It is routinely obtained from the spatial correlation function of the hexatic order parameter,
\begin{equation}
  \Psi_{6} = \biggl\langle \frac{1}{N} \sum_{j=1}^{N} \psi_{6}^j \biggr\rangle,
\label{Psi_6_eq}
\end{equation}
\begin{equation}
  \text{where}\quad \psi_{6}^j = \frac{1}{n_{j}}\sum_{k=1}^{n_{j}}\text{exp}(i6\theta_{jk}), 
\end{equation}
\begin{figure*}[h]
\includegraphics[width=0.85\textwidth,keepaspectratio]{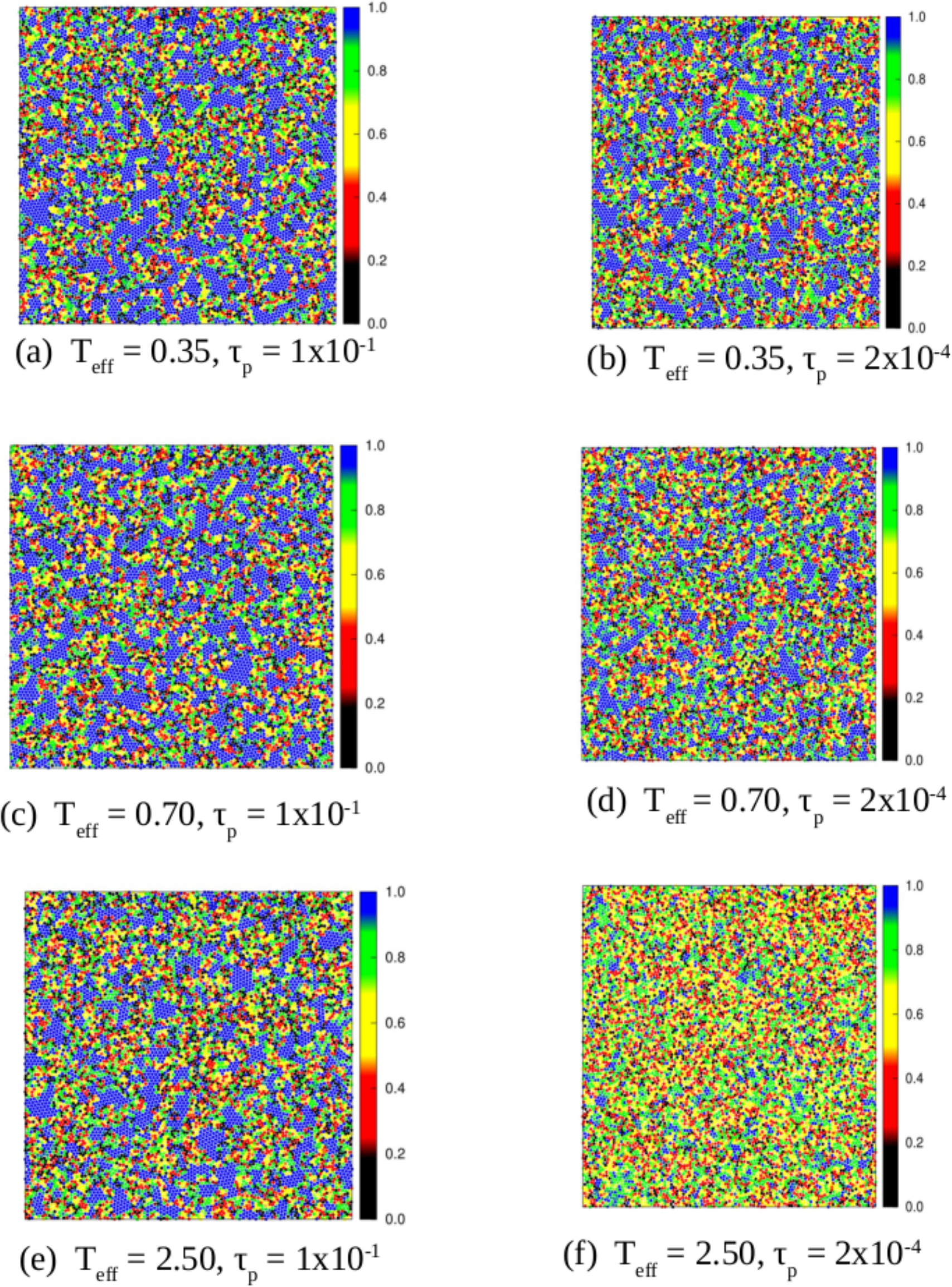}
\caption{Images show color-maps of the local bond orientation order parameter $\psi_{6}^j$, where $j$ is any particle in the system. At lower effective temperatures, increasing the persistence time brings only a marginal increase in the local bond order.}
\label{psi6-plot}
\end{figure*}
and  $\theta_{jk}$ is the angle between the separation vector $\bm{r}_{jk}$ and a reference axis, say the $\hat{x}$ axis.  Here $n_{j}$ refers to the number of Voronoi (nearest) neighbors of any particle particle $j$ in the liquid, and $\langle \cdots \rangle$ describes the ensemble average over statistically independent snapshots taken in the steady state. By mathematical construction, $\psi_{6}^{j}$ can range anywhere in the interval [0,1] with the values 0 and 1 corresponding respectively to a locally disordered and a locally ordered configuration. In Figure \ref{psi6-plot}, we show instantaneous snapshots of our local order parameter $\psi_{6}^j$ to visualise the presence of ordered and disordered regions in our system. It is evident from the figure that the degree of local bond orientation ordering tends to increase only marginally with the persistence time, indicating that the hexatic length scale may not be the dominant length scale that signifies sluggish dynamics.  To quantitatively extract a hexatic length scale $\xi_6$, we use the spatial correlation of hexatic order parameter,
\begin{equation}
g_{6}(r)=\frac{L^{2}}{2\pi r\Delta r N(N-1)}\sum_{j \neq k} \delta (r-|\bm r_{jk}|)\psi_{6}^{j}\psi_{6}^{k*}. 
\label{g6_eq}
\end{equation}
A numerical fit of the peak values of $g_6(r)/g(r)$ data to the Ornstein-Zernike (OZ) correlation function 
\begin{equation}
  g_{6}(r)/g(r) \sim r^{-1/4}\text{exp}(-r/\xi_{6})
\label{g6-OZ-fit}
\end{equation}
directly yields the hexatic length scale $\xi_6$ \cite{tanaka2010critical}. In Figure \ref{g6-fit}, we a plot of this hexatic correlation function normalized to the radial distribution function $g_6(r)/g(r)$, at a fixed persistence time $\tau_p$. 
The solid lines are a fit to the OZ prediction \ref{g6-OZ-fit} to extract the hexatic length scale $\xi_6$. It is clear, at least qualitatively, that $\xi_6$ tends to increase as the effective temperature $T_{\text{eff}}$ is lowered. To further illustrate the dependence on the persistence time $\tau_p$, we show in Fig. \ref{xi6-vs-T}, a plot of $\xi_6$ vs. $T_{\text{eff}}$, at various values of persistence times. It is evident from this figure that at any temperature, the effect $\tau_p$ on $\xi_6$ grows is only marginal. These observations are consistent with our data on local order, shown earlier in Fig. \ref{psi6-plot}. Next, we will discuss another length scale that arises from the microscopic analysis of particle diffusion.

\begin{figure}[h]
\includegraphics[width=0.85\textwidth,keepaspectratio]{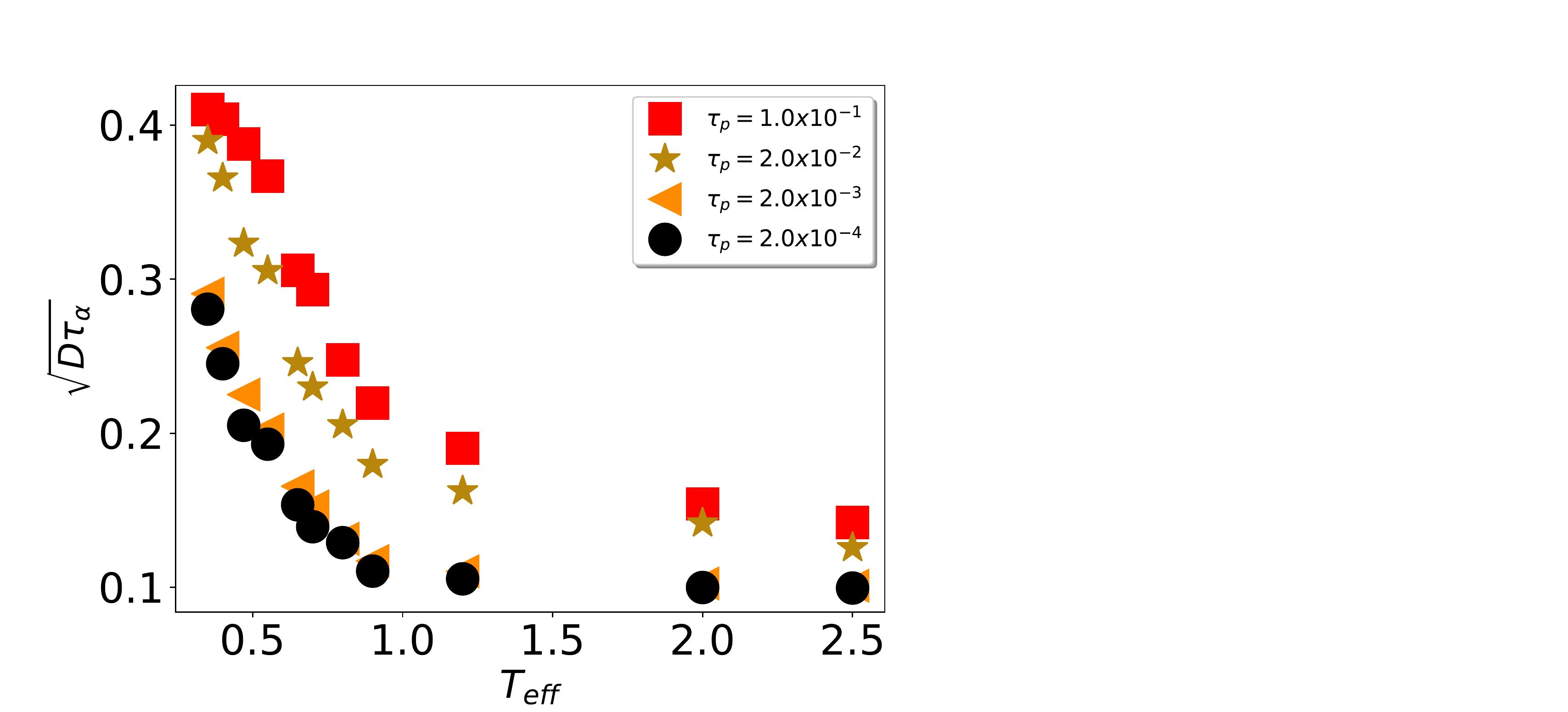}
\caption{Length scale ($\sqrt{D\tau_\alpha}$) obtained from diffusion coefficient increases with inverse of $T_{\text{eff}}$ for various persistence time.}
\label{Dtau-vs-T}
\end{figure}

\subsection{Diffusion Length Scale $\sqrt{D \tau_{\alpha}}$}
\begin{figure}[h]
\includegraphics[width=0.85\textwidth,keepaspectratio]{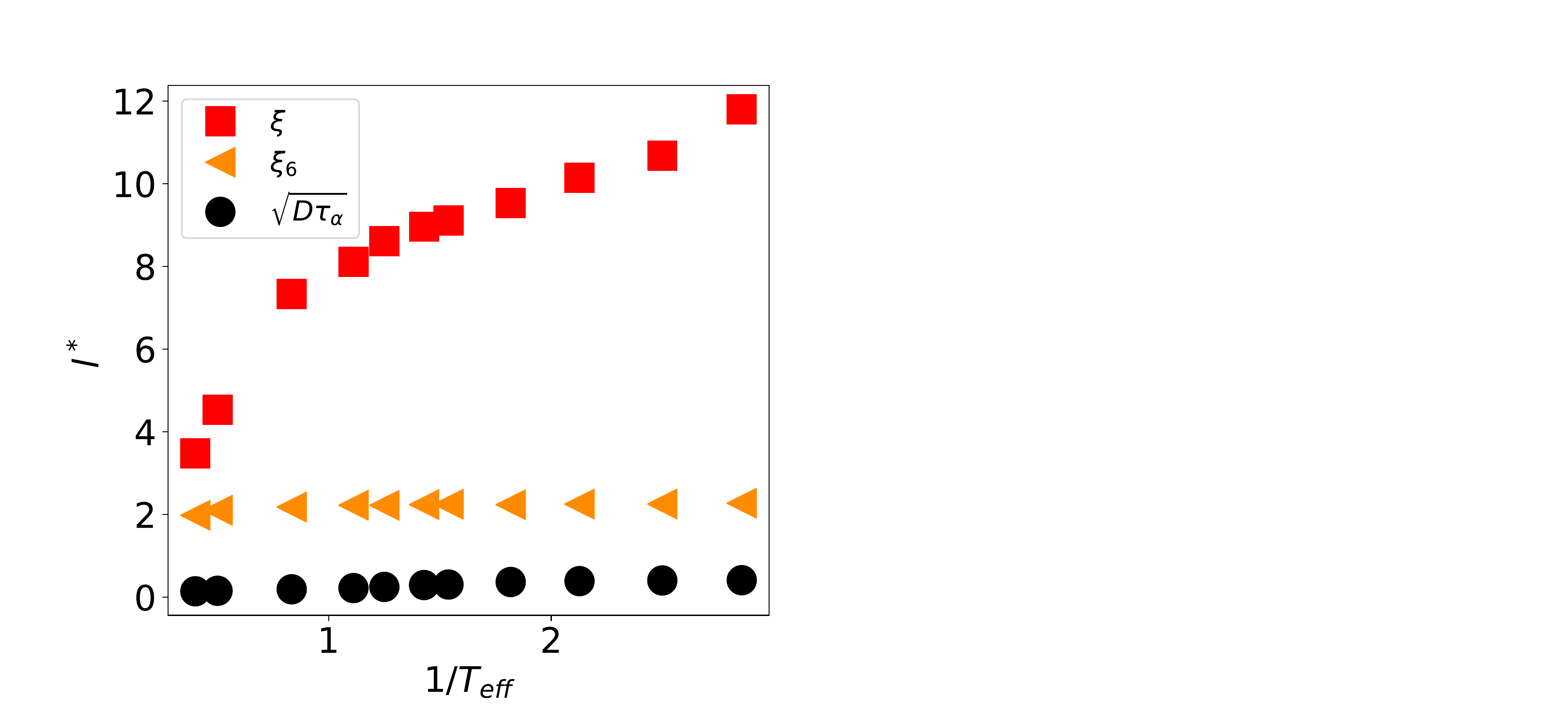}
\includegraphics[width=0.85\textwidth,keepaspectratio]{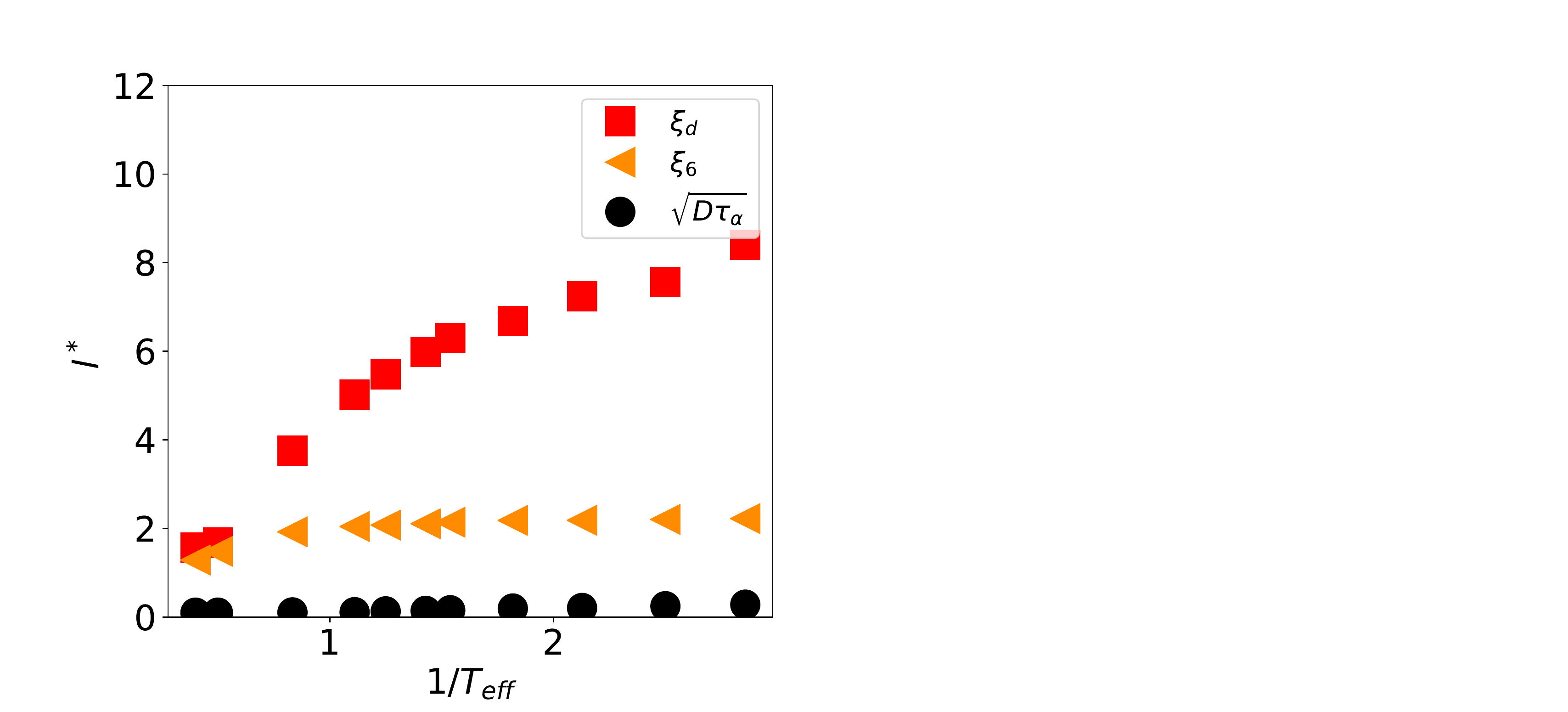}
\caption{Length scales vs. effective temperature for two different persistence
times, namely for $\tau_p = 0.1$ (top), and $\tau_p = 0.0002$ (bottom). It is
evident that all length scales grow with inverse effective temperature. Note
that for any given persistence time, the dynamical length scale $\xi$
progressively decouples from the other length scales as the effective
temperature is lowered.  The onset of this decoupling however, is pushed to
higher effective temperatures as the persistence time is increased.}
\label{all-lengths}
\end{figure}

A  microscopic length scale can be constructed from the diffusion coefficient $D$ and the $\alpha$-relaxation time $\tau_\alpha$ in the form of $\sqrt{D\tau_{\alpha}}$ \cite{flenner2011analysis}. The realization of this as a growing length scale can be argued from the fact $\tau_\alpha$ is to a good approximation of the onset of the Fickian diffusion in typical liquids, which clearly becomes larger as the liquid is taken to lower temperatures shown in Fig. \ref{Dtau-vs-T}. In Figure \ref{all-lengths}, we plot all these length scales as a function of temperatures for two different values of persistence times. Below we discuss our observations on the dependence of the length scales reported in this paper, namely $\xi, \xi_6$ and $\sqrt{D \tau_\alpha}$ as a function of the state parameters, $\tau_p$ and $T_{\text{eff}}$. 
\subsection{Decoupling of Length Scales on $T_{\text{eff}}$ and $\tau_p$} It is
evident from the figure \ref{all-lengths} that all length scales tend to grow
with inverse effective temperature but the dynamic length scale $\xi$ clearly
dominates and is therefore decoupled from the others. This is especially true
at lower effective temperatures and for all the persistence times considered here.
The role of persistence time is to merely push the onset of decoupling towards
higher temperatures.  One can observe this qualitatively in Fig.
\ref{all-lengths} where we show our data on all the length scales for two
extreme values of $\tau_p$, namely 0.0002 (bottom) and 0.1 (top).  Similar
decoupling of length scales have been studied in volume fraction dependent
studies of passive liquids maintained at fixed temperature
\cite{kawasaki2011structural}. We summarize below to conclude our findings.

\section{\label{summary} Summary}
To summarize our paper, we have performed extensive numerical simulations to extract various length scales near the glass transition in a model ``athermal'' active glass forming liquid. The dependence of these length scales on the effective temperature and persistence times is carefully explored, and a detailed comparison between the dynamic and static length scales is also presented. An important finding of our work is a simple exponential scaling law, $\tau_\alpha \sim \text{exp}(\mu \xi/T_{\text{eff}})$, that connects the dynamic length scale $\xi$ to the concomitant growing time scale $\tau_\alpha$ through the concept of a chemical potential $\mu$ - a budget that needs to be scaled in order to form clusters of slow moving particles. The quantity $\mu \xi$ then refers to a growing free energy barrier that can account for slower re-arrangement of the aforementioned clusters, as $T_{\text{eff}}$ is reduced. Our estimated $\xi$ also agrees very well with the size of these clusters, further asserting our claim. We also observe that $\xi$ dominates over the static length scale $\xi_6$ at all temperatures and persistence times considered here. The latter is saturated at lower temperatures and is only  marginally affected by the persistence time in this range. There is therefore no general correspondence between $\xi_{6}$ and $\xi$ in the case of an ``athermal'' active liquid. The findings reported in our paper are robust over three orders of magnitude variation in persistence time and hence should be of great utility to researchers interested in the glassy dynamics of a generic active liquid. 

\section{Acknowledgements}
We thank Ethayaraja Mani and Smarajit Karmakar for comments and discussions. All simulations were done on the VIRGO super cluster of IIT Madras and the HPC-Physics cluster of our group. Support from
DST INSPIRE Faculty Grant 2013/PH-59 and the New Faculty Seed Grant (NFSG), IIT Madras, is gratefully acknowledged.
\bibliography{manuscript}
\end{document}